\documentstyle[epsf]{mn}

\title[Bayesian SZE detection]{A Bayesian non-parametric method to detect clusters   
in Planck data.}  
\author[J.M Diego et al.]  
   { J.M  Diego$^1$, P. Vielva $^{2,3}$, E. Mart{\'\i}nez-Gonz{\'a}lez$^2$,   
     J. Silk$^1$ and J.L. Sanz$^2$. \\  
   $^1$Department of Astrophysics. Univerity of Oxford. 1, Denys Wilkinson Building,  
       Keble Road,  Oxford OX1 3RH,  UK\\   
   $^2$Instituto de F{\'\i}sica de Cantabria (CSIC - UC) Avda. Los  
     Castros s/n, 39005 Santander, Spain\\   
   $^3$Departamento de F{\'\i}sica Moderna. Universidad de Cantabria,   
     Avda. Los Castros s/n, 39005 Santander, Spain\\}

\pagerange{\pageref{firstpage}--\pageref{lastpage}}  
  
\begin{document}  
  
\maketitle  
  
\label{firstpage}  
\begin{abstract}  
  
We show how one may expect  a significant number   
of SZ detections in future Planck data without any   
of the typical assumptions needed in present component separation  
methods, such as  about the power spectrum   
or the frequency dependence of any  of the components,   
circular symmetry or a typical scale for the clusters.   
We reduce the background by subtracting an estimate of the   
point sources, dust and CMB. The final SZE map is estimated in Fourier space.   
The catalogue of returned clusters is complete above   
flux $\approx 200$ mJy (353 GHz)     
while the lowest flux reached by our  method is $\approx 70$ mJy (353 GHz).   
We  predict a large  number of detections ($\sim 9000$) in 4/5 of   
the sky.   
This large number of SZ detections will allow a robust and consistent   
analysis of the evolution of the cluster population with redshift and    
will have important implications for determining the best   
cosmological model.   
  
\end{abstract}  
  
  
\section{Introduction}\label{section_introduction}  
  
The distortion in the radiation intensity of CMB photons produced when they   
traverse  the hot intracluster plasma in the direction of a galaxy cluster   
(Sunyaev-Zel'dovich effect, Sunyaev \& Zel'dovich 1972)   
is one of the most promising effects  being  studied   
for exploring cosmological models.  
In recent years, several groups have been working on the detection   
of the SZE and many detections have been reported (e.g. Birkinshaw et al. 1984,   
Carlstrom et al. 1996, Pointecouteau et al. 2001). The SZE is growing in interest   
as the number of experiments and their quality is increasing.   
The number of experiments devoted to these kind of observations, as well   
as their unprecedented quality, will allow a variety of analyses which,   
combining the SZ data with other data or alone, could be applied to the   
study of the  intracluster media, its origin and evolution;   
the abundance of galaxy clusters and its important cosmological   
implications; the determination of   
the cosmological distances to the most distant galaxy clusters, etc.  
One of these experiments is the approved Planck satellite (scheduled launch   
in 2007). This satellite will observe the full sky at mm frequencies   
(30 GHz $< \nu <$ 857 GHz) and with resolutions ranging from $FWHM = 30$   
arcmin to $FWHM = 5$ arcmin. Previous studies have shown that this satellite   
will produce a full sky cluster catalogue with about $5000-50000$   
clusters; the final number depending on the cosmological model, the   
Planck effective sensitivity and the method used to identify the different   
component contributions. This paper will be focused on this  last point.  
  
Recent proposed component separation methods,   
Wiener filter (WF, Tegmark \& Efstathiou 1996, Bouchet et al. 1997),     
maximum entropy (MEM, Hobson et al. 1998, 1999),   
fast independent component analysis (FastICA, Maino et al. 2001),   
mexican hat wavelet analysis (MHW, Cay{\'o}n et al. 2000, Vielva et al. 2001a),   
and adaptive filter analysis (AFA, Tegmark \& de Oliveira-Costa 1998,  
Sanz et al. 2001),  
are being tested in order to define a well established   
method to perform the component separation on the Planck data.   
However, it will be extremely difficult to define the best method   
since some methods will work better than others under certain   
circumstances,  and it will not be surprising if, at the end,   
the final component separation method results in a   
combination of a variety of methods (e.g. MEM $+$ MHW, Vielva et al. 2001b).   
  
Some methods try to separate all the components   
simultaneously. To do this in an effective way, some {\it a priori}   
information is needed. Commonly, the power spectrum of several (if not all)  
components and the frequency dependence of the components must be given   
(WF, MEM). Another typical assumption is that all the components are independent and   
non-Gaussian except maybe one, the CMB (FastICA). \\   
In the case that the assumed information is close enough to reality, all of these methods   
work very well.   
On the other hand, if the {\it a priori} information is wrong,   
the result of the component separation will be biased with respect   
to the underlying real signal. This could have important consequences   
on the analysis of the final data maps. One of the risks in the   
{\it simultaneous all component separation methods} is then that an error   
in the estimation of one of the components must be compensated by an error   
in one (if not all) of the other components due to the constraint that   
the sum of all the recovered components must equal the data.  
  
This problem can be partially reduced by using {\it single   
component separation methods} like the MHW or AFA which have been successfully   
applied to the separation of point sources (Cay{\'o}n et al. 2000,   
Vielva et al. 2001a) and the SZ effect (Herranz et al. 2002a,b). 
These methods have the advantage, over the previous ones, that they do not   
need to assume anything about the Galactic components or the CMB.  
The information they need is taken  
directly from the data (the power spectrum of the   
background and the beam shape).   
The only thing they have to assume is a scale    
and the circular symmetry of the source. In the MHW technique applied  
to detect point  sources, the optimal scale can be  
obtained from the background and the beam   
scale. In this sense, the analysis of the point sources based on the  
MHW technique  
is very robust since all the assumptions are taken from the data.   
When the AFA is applied to the detection of the SZ effect, a prior knowledge   
of the scale and shape (asymmetry)  of the clusters is needed (Sanz et al. 2001).   
This problem can be overcome by applying the filter at different   
scales (Herranz et al. 2002a,b). The problem of asymmetry   
in the resolved clusters can only be solved by rotating an axis-asymmetric   
filter which will reduce significantly the speed of the algorithm.   
  
In this work we propose an alternative method which can be applied to the   
detection of the SZ signal on the future Planck data.   
The main points of our method are the following:\\  
  
\noindent  
$\bullet$ We consider a Bayesian and non-parametric method without prior  
	  knowledge about the power spectrum of any   
          of the components. We will show how the method allows us to include   
          information about the power spectrum of the SZE component.   
          However the final result will not depend significantly on the   
          particular choice of this power spectrum and arbitrary power   
          spectra can be considered provided they obey some general rules. \\  
  
\noindent  
$\bullet$ The method is easy to implement and fast since it is a   
          non-iterative method and all the equations are solved in   
          Fourier space mode by mode.\\

\noindent  
$\bullet$ We do not need any prior knowledge about the frequency dependence   
          of the components other than the SZ, and, obviously, the CMB.   
          We only require that we have at least one channel which is clearly dominated   
          by dust. However, results from BOOMERANG and IRAS suggest that the 857 GHz channel   
          in Planck will be dominated by dust.\\  
  
\noindent  
$\bullet$ The method works for any kind of shapes and sizes of the   
          galaxy clusters.\\  
  
\noindent  
$\bullet$ The method uses all the information available (all the channels). \\

As mentioned in the third point, we will assume only  knowledge of   
the frequency dependence of the SZ effect (see fig. \ref{fig_fx}).   
This is a well established assumption   
as the physic of the SZ effect is very well known. We would like to   
remark that, although in this work we will assume only the non-relativistic   
corrections, our conclusions could be extended to include   
the relativistic corrections (provided the temperature of the cluster is known).   
We also did not consider the kinetic contribution to the SZ effect   
since it is of order 30 times smaller than   
the thermal part. However, this component could be estimated (in some   
clusters) as the residuals after the thermal contribution has been determined.\\  
  
The structure of the paper is the following.   
In section \ref{data_set} we present the Planck simulations  
that have been used to test the method. We describe the method  
and the way it is implemented in section \ref{section_method}.  
We apply the method to the {\it realistic} Planck simulations     
and show the results in section \ref{section_application}.   
Finally, we compare briefly with other methods  
in section \ref{section_discussion}.   
The possibilities of the recovered SZE map are also highlighted in this   
section.

\section{Data set: realistic Planck simulations}  
\label{data_set}  
  
\begin{figure}  
   \begin{center}  
   \epsfysize=18.cm   
   \caption{Initial simulated maps at the Planck frequency channels. They  
   contain: CMB, thermal Sunyaev-Zel'dovich effect, thermal dust  
   emission, free-free, synchrotron, spinning dust, point source  
   emission and instrumental Gaussian white noise. In this plot and   
   in the next ones, the color table is in units of $\Delta T/T$.}  
   \label{fig_10chan}  
   \end{center}  
\end{figure}  

In order to check the power of the method, we have performed   
{\it realistic} Planck simulations. The simulations are realistic   
in the sense that they include all the main features of the   
Planck satellite such as the corresponding noise level in each channel,   
pixel size and antenna beam (see table 1).  
They are also realistic in the sense that all the components   
(CMB, Galactic components, extra-galactic point sources and SZ)   
were simulated including the latest information we have   
about these components. The simulations were done   
in patches of the sky of $12.8^{\circ}\times12.8^{\circ}$   
although the method can be easily extended to include all the sphere.  
For the sake of simplicity, we will not include the effect of the bandwidths   
in our simulations although there is no problem if the bandwidths have to be included.   
Therefore, we will simulate the different maps only at the central frequency of the   
Planck channels. \\  
  
The CMB simulation has been done for a spatially flat  
$\Lambda$CDM Universe with  
$\Omega_m = 0.3$ and $\Omega_{\Lambda} = 0.7$,  
using the $C_l$'s generated with the CMBFAST  
code (Seljak \& Zaldarriaga, 1996). It is a Gaussian realization.  
  
The thermal Sunyaev-Zel'dovich (SZ) effect simulation was made   
for the same cosmological model. The cluster population was modeled   
using Press-Schechter (Press \& Schechter 1974)   
with a Poissonian distribution in the angular coordinates of the 2D map,   
$\theta$ and $\phi$.  
The model was selected by fitting the cluster population as a function of   
$z$ to several X-ray and optical cluster data sets. In that fit we obtained certain   
values for the cosmological parameters as well as an estimate   
for the parameters involved in the cluster scaling relations   
$T-M$ and $L_x-T$ (see Diego et al. 2001a for a discussion).  
  
The extra-galactic point source simulation was    
performed following the model of Toffolatti et al. 1998  
assuming the cosmological model  
indicated above. The simulation include radio flat-spectrum  
and infrared sources. VLA and IRAS catalogues were   
used to fix the model. The predictions obtained with this  
model are compatible with ISO and SCUBA data (see the above  
paper for more details).

\begin{table*} \label{table1}  
   \begin{center}  
         \begin{tabular}{|c|c|c|c|c|c|c|c|c|c|c|}  
	 \hline  
	 Frequency & FWHM & Pixel size & $\sigma_{noise}$ &  
         CMB  &  TDust  & FF & Synch. & SDust & PS & SZ \\  
	 (GHz) & (arcmin) & (arcmin) & $(10^{-6})$ &  
         $(10^{-6})$ & $(10^{-6})$ & $(10^{-6})$ & $(10^{-6})$ &  
         $(10^{-6})$ & $(10^{-6})$ & $(10^{-6})$ \\  
	 \hline  
	 857 & 5.0 & 1.5 & 22211.10 &  
	 42.70 & 140000.00 & 39.60 & 17.30 & 0.00 & 11400.00 & 19.60 \\  
	 \hline  
	 545 & 5.0 & 1.5 & 489.51 &  
         42.70 & 1090.00 & 1.10 & 0.58 & 0.00 & 92.80 & 9.86 \\  
	 \hline  
	 353 & 5.0 & 1.5 & 47.95 &  
         42.70 & 58.50 & 0.23 & 0.15 & 0.00 & 5.16 & 3.93 \\  
	 \hline  
	 217 & 5.5 & 1.5 & 15.78 &  
         42.50 & 7.53 & 0.16 & 0.12 & 0.00 & 1.57 & 0.03 \\  
	 \hline  
	 143 & 8.0 & 1.5 & 10.66 &  
         41.0 & 2.33 & 0.21 & 0.20 & 0.00 & 1.91 & 1.40 \\  
	 \hline  
	 100 (HFI) & 10.7 & 3.0 & 6.07 &  
 	 39.40 & 1.07 & 0.36 & 0.39 & 0.00 & 2.90 & 1.65 \\  
	 \hline  
	 100 (LFI) & 10.0 & 3.0 & 14.32 &  
         39.8 & 1.06 & 0.35 & 0.39 & 0.00 & 3.11 & 1.73 \\  
	 \hline  
	 70 & 14.0 & 3.0 & 16.81 &  
         37.6 & 0.54 & 0.67 & 0.87 & 0.27 & 4.00 & 1.59 \\  
	 \hline  
	 44 & 23.0 & 6.0 & 6.79 &  
         33.2 & 0.24 & 1.64 & 2.65 & 3.17 & 5.82 & 1.16 \\  
	 \hline  
	 30 & 33.0 & 6.0 & 8.80 &  
         29.3 & 0.12 & 3.56 & 6.87 & 8.94 & 8.35 & 0.89 \\  
	 \hline  
      \end{tabular}  
      \caption{Experimental constrains and simulation characteristics  
      at the 10 Planck channels. The  
      antenna FWHM is given in column 2 for the different frequencies  
      (a Gaussian pattern is assumed in the HFI and LFI  
      channels). Characteristic pixel sizes are shown in column 3. The  
      fourth column contains information about the instrumental noise  
      level, in $\Delta T/T$ per pixel. In columns 5 to 11 we show  
      the dispersion of the simulated components (CMB, thermal dust  
      emission, free-free, synchrotron, spinning dust, point sources  
      and Sunyaev-Zel'dovich effect respectively) in  $\Delta T/T$   
      per pixel, after beam convolution.}  
    \end{center}  
\end{table*}  
  
We have simulated four different Galactic emission sources:  
thermal dust, free-free, synchrotron and spinning dust.  
  
The thermal dust emission was simulated using the data and the model  
provided by Finkbeiner et al. (1999). This model assumes that  
dust emission is due to two grey bodies: a \emph{hot} one with a  
dust temperature of ${T_D}^{hot} \simeq 16.2K$ and an  
emissivity $\alpha^{hot} \simeq 2.70$, and a \emph{cold} one with a  
${T_D}^{cold} \simeq 9.4K$, and an $\alpha^{cold} \simeq 1.67$.  
These quantities are mean values. The temperatures and emissivities  
change from point to point. An exhaustive description of the   
model is given in their paper, where the authors combined    
data from DIRBE, IRAS and FIRAS to fit the dust emission at high   
frequencies (500 GHz to 3000 GHz). We will use their best fitting   
model in this work.   
  
The distribution of free-free emission is poorly known.  
Present experiments such as the H-$\alpha$ Sky Survey  
\footnote{http://www.swarthmore.edu/Home/News/Astronomy/}  
and the WHAM project \footnote{http://www.astro.wisc.edu/wham/}  
will provide maps  
of $H_{\alpha}$ emission that could be used as a template.  
In this work, we have created the free-free template correlated    
with the dust emission as proposed in Bouchet, Gispert \& Puget (1996).   
The frequency dependence of the  
free-free emission is assumed to change as $I_{\nu} \propto  
{\nu}^{-0.16}$, and is normalized to give an RMS  
temperature fluctuation of $6.2 \mu K$ at 53 GHz.  
  
Synchrotron emission simulations have been done using the  
all sky template provided by P. Fosalba and G. Giardino  
in the FTP site: \emph{ftp://astro.estec.esa.nl}. This map  
is an extrapolation of the 408 MHz radio map of Haslam et al.  
1982, from the original $1^{\circ}$ resolution to a resolution  
of about $5$ arcmin.   
The additional small-scale structure  
is assumed to have a power-law power spectrum   
with an exponent of $-3$.   
We have done an additional extrapolation to the smallest   
scale ($1.5$ arcmin) with the same power-law.  
We also include in our simulations the information on the changes of   
spectral index as a function of electron density in the Galaxy.  
This template has been done combining the Haslam  
map with the Jonas et al. 1998 at 2326 MHz and with the Reich  
\& Reich 1986 map at 1420 MHz, and can be found in the previous  
FTP site.  
  
We have also taken into account possible galactic emission  
due to spinning grains of dust, proposed by Draine  
\& Lazarian 1998. This component could be  
important at the lowest frequencies of the Planck channels  
(30 and 44 GHz) in the outskirts of the galactic plane.   
  
  

\section{A method in 2 steps}\label{section_method}  
  
In this section, we describe our proposed new method to estimate the SZ   
thermal contribution to the mm data in the 10 Planck channels.   
A detailed description of the mission can be found in the official   
Planck web address http://astro.estec.esa.nl/Planck/.  
  
\begin{figure}  
   \begin{center}  
   \epsfxsize=8.cm   
   \begin{minipage}{\epsfxsize}\epsffile{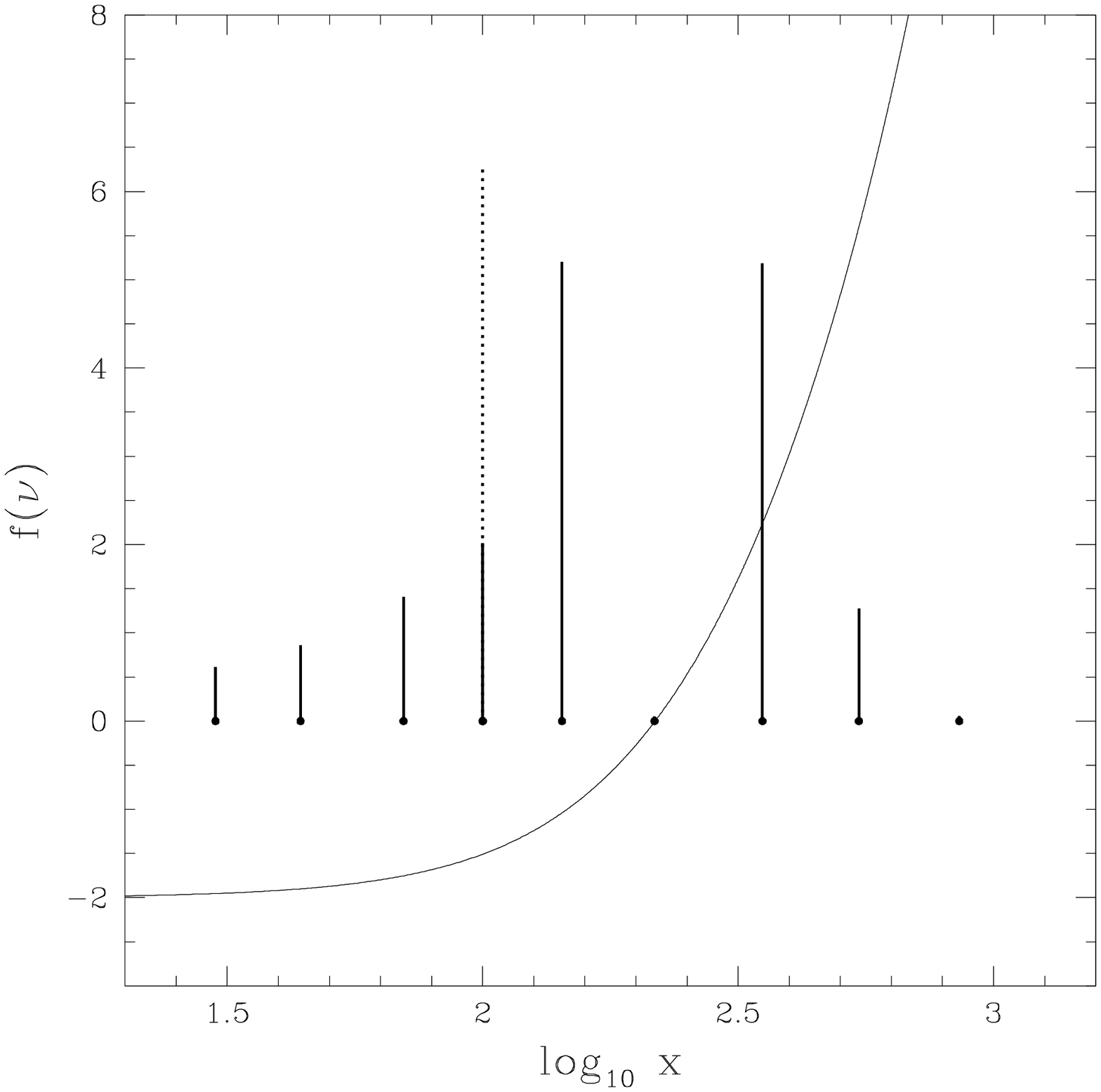}\end{minipage}  
   \caption{\label{fig_fx}  
            $f(\nu)$ factor appearing in eqn. (\ref{eq_SZ}). The vertical   
            lines are proportional to the signal-to-noise ratio for the SZE   
            expected in each one of the Planck channels.   
            Planck frequencies are centered at   
            30, 44, 70, 100 (LFI \& HFI), 143, 217, 353, 545 and 857 GHz.   
            The planned bandwidths for Planck (not considered in this work)   
            are $\delta\nu/\nu = 0.2$ for the LFI (30-100 GHz)   
            and $\delta\nu/\nu = 0.25$ for the HFI (100-857 GHz).   
            The channels at 217 and 857 GHz show a tiny S/N ratio.   
            At 100 GHz there are two channels; LFI channel (solid line) and   
            HFI channel (dotted line).   
            }  
   \end{center}  
\end{figure}

The expected sensitivity of Planck to detect the SZ effect in each one of its   
10 channels is shown in Fig. \ref{fig_fx} as vertical lines centered in each one   
of the central frequencies. The amplitude of this lines is proportional   
to $f(\nu)/\sigma(\nu)$ where $\sigma(\nu)$ is the sensitivity per resolution element   
of the channel at frequency $\nu$.  
The factor $f(\nu)$ is just the frequency dependence of the thermal SZ effect:   
\begin{equation}  
\frac{\Delta T}{T}(\nu) = f(\nu)y_c  
\label{eq_SZ}  
\end{equation}  
where $y_c$ is the Compton parameter, $T$ is the thermodynamic  
mean temperature of the CMB ($T \approx 2.73$ K)   
and $\Delta T$ is the change in the thermodynamic CMB temperature induced by the SZE.   
The thermodynamic temperature is related to the intensity, $I$,  through:  
\begin{equation}  
T(\mu K) \approx \frac{1}{24.8}{\Big({\frac{\sinh{(x/2)}}{x^2}}\Big)^2}I(Jy~sr^{-1})  
\end{equation}  
where $x \approx \nu/56.8$ GHz.\\  
>From figure \ref{fig_fx} it can be seen that the best channels are   
those between 100 and 353 GHz. Although the channel at 217 GHz does not   
seem to be relevant, it is in fact one of the most important to detect the SZ   
effect since at this frequency the thermal SZ effect is expected to be   
negligible. The other channel that does not seem to be relevant is the   
highest one   
at 857 GHz which is expected to be completely dominated by the dust emission   
coming from our Galaxy (see Table 1).   
However, as we will see later, both channels   
will play a crucial role in our method. \\  
The method is in fact divided in two main steps.\\

\noindent  
In the first step -\emph{map cleaning}- we reduce the contribution of   
certain components (point sources, dust and CMB) under the assumption   
that point sources are unresolved, the thermal dust emission is,  
at different frequencies,   
the same spatial pattern times a parameter which depends on the frequency   
and the CMB is frequency-independent.   
This process will increase the noise level of the maps but will increase as   
well the S/N ratio of the SZE signal. \\  
  
  
   
\noindent  
In the second step -\emph{Bayesian approach}-  
we develop a  
method to search for the Compton parameter in each pixel  
responsible of the SZ signature in our {\it clean} maps.   
We will define our   
approach in terms of Bayes' theorem.

\subsection{Map cleaning}  
\label{cleaning}  
A typical CMB experiment will measure not only the CMB signal   
but also other additional components such as the Galaxy (synchrotron   
and free-free emission at low frequencies and dust emission at high   
frequencies), extra-galactic sources which for the Planck resolution   
will appear as unresolved point sources, and finally the SZE.   
The integrated contribution of these components is detected with an   
antenna having a given response (different at each channel).   
In addition we have to include   
the noise of our detectors which also will depend on the frequency (channel).   
The final signal at a given frequency will be therefore:   
\begin{equation}  
\frac{\Delta T}{T}(\nu, \vec{x}) =   
A\otimes\sum_i \frac{\Delta T_i}{T}(\nu, \vec{x}) +  
\frac{\Delta T(\nu, \vec{x})^N}{T}  
\label{eqn_DeltaT}  
\end{equation}  
where $\sum_i$ is a sum over all the components (CMB, Galactic components,   
extra-galactic point sources and SZE) and  
$\Delta T(\nu, \vec{x})^N/T$ contains the contribution   
due to the noise in the receivers. We have considered a white Gaussian uniform noise.  
In a real situation, the noise will not be uniformly distributed.   
This effect is minimized when taking small sky patches. \\

$\Delta T/T(\nu, \vec{x})$ denotes the measured temperature of the sky minus   
the temperature of the CMB ($T_{CMB} \approx 2.73$ K) divided by $T_{CMB}$.  
The term $A\otimes$ denotes the convolution with the antenna. There should be   
an additional term in the previous equation to account for the frequency   
response of our experiment which is not a delta function at the frequency   
$\nu$. Therefore a real experiment will measure  
$\Delta T/T(\nu_1 - \nu_2, \vec{x})$.   
This is just another convolution of Eq. \ref{eqn_DeltaT} with the frequency   
response of the instrument centered at the frequency $\nu$. For simplicity we will   
not consider the bandwidth in our calculations although it could easily be included.\\      
By looking at Eq. \ref{eqn_DeltaT}, it is easy to   
understand the complexity of the component separation problem.   
In this work we are only interested in one of these components, the SZE.    
The complexity of estimating that component could be reduced if we   
can subtract first, or at least reduce significantly, the contribution   
of some of the other components in Eq. \ref{eqn_DeltaT}.  
By reducing the contribution of some of the dominant components,  
the S/N ratio of the SZ signal can be increased, since the smaller   
the background the better our determination of the signal. But one should   
be careful in the process of subtracting some of the other components   
since we do not want to remove any SZE signal. \\  
There are several components which can be easily subtracted   
from the Planck data (or at least reduce their contribution to the   
background) without subtracting any significant thermal SZE signal. \\

\noindent  
{\bf Point sources.}  
  
\noindent  
The point source contribution is expected to be specially  
relevant at the highest Planck  
frequency channels. The point source emission at these high   
frequencies is due to infrared sources. At the lowest Planck  
frequencies, the point source emission is mainly due to radio   
flat-spectrum AGNs. The knowledge of the point source emission  
at the intermediate Planck channels is really poor.  
In fact, the determination of the point source emission at  
these frequencies is one of the challenges of the Planck mission.  
  
The detection of the point source emission is a special issue  
for the component separation problem. There are two main differences  
between this emission and the other foregrounds.  
First, the frequency behaviour can change significantly from one point source   
to another. Second, the point source emission  
has a typical scale: the beam width.  
These properties suggest that common component separation   
methods such as MEM, WF or neural networks   
are not the best techniques to detect the point source emission.  
  
We have applied the MHW technique first  
described in Cay{\'o}n et al. (2000) and later extended in  
Vielva et al. (2001a).  
Wavelets are a powerful tool to detect point sources.   
When a signal with a characteristic scale is analyzed with  
a wavelet adapted to its shape, its wavelet coefficients are   
amplified with respect to the background coefficients. This   
amplification occurs at scales around the characteristic scale (Cay{\'o}n et al. 2000).   
We can increase the number of detections   
by looking at the scale which maximizes the amplification  
(Vielva et al. 2001a). This optimal scale can be determined   
directly from the data. Therefore, there is no need for    
assumptions either about the nature or characteristics of the  
underlying signals (e.g. spectral behaviors, power spectra,  
pdf, etc).   
The optimal pseudo-filter for the detection of point sources    
which are convolved with a Gaussian beam (at least for 2D images with  
a power spectrum described by $C_l\sim l^{-2}$  
around the point source scale) is the MHW (Sanz et al. 2001).  
A detailed description of the point source detection  
algorithm can be found in Vielva et al. (2001a). Here we will just describe   
the main steps. \\  
First, we search for the optimal MHW scale that  
maximizes the amplification (source amplitude ratio, in dispersion  
units, between wavelet and real spaces) at each frequency. As we mentioned  
above, all the information we need is in the data itself. Basically,  
the function that must be maximized is:  
\begin{equation}  
\textrm{$\mathcal{A}$}(R) \propto \int{dk kP(k) |\widehat{\psi}(Rk)|^2}  
\end{equation}  
where $P(k)$ is the power spectrum of the map,   
$|\widehat{\psi}(Rk)|$ is the Fourier transform of the MHW and R is  
the MHW scale. Once the optimal scale is determined, we select   
certain point source candidates (those maxima at the MHW optimal scale map that  
are above a certain level). The next step is to get the amplitude of  
those point source candidates. The amplitude estimation is done by fitting the  
theoretical dependence of the point source wavelet coefficient with the MHW  
scale. Then, we convolve the map with three additional MHW (at  
adjacent scales to the optimal one) and we fit the obtained  
coefficients to the theoretical curve:  
\begin{equation}\label{eq:Coeff}  
   \frac{w(R)}{R} = 2\sqrt{2\pi}B  
                              \frac{(R/\sigma_a)^2}  
                              {(1 + (R/\sigma_a)^2)^2}  
\end{equation}  
where $w(R)$ is the wavelet coefficient at the scale $R$, $\sigma_a$ is  
the beam dispersion and $B$ is the PS amplitude (the parameter to be  
determined). After this, a detection criterion is applied to select  
the real point sources (see Vielva et al. 2001a) and those point sources are  
subtracted from the original data.\\

\noindent  
{\bf Dust.}  
  
\noindent  
The second contribution which can be easily subtracted is the thermal  
dust emission.   
This emission can be important at frequencies above   
$\approx 100$ GHz. At frequencies of $\approx$ 350 GHz, dust starts to dominate    
over the other components and its contribution is even more important at higher   
frequencies. The Planck channel at 857 GHz is expected to be completely   
dominated by dust (as suggested by the BOOMERANG and IRAS results).   
This map can therefore be used as a spatial pattern   
of the distribution of dust in our Galaxy. The only thing we need to know    
to subtract the dust from the other channels, is the frequency   
dependence of the dust.   
However, since we do not want to assume any frequency dependence (and this   
is unknown in the range of frequencies of Planck) we need   
to look for other alternatives to subtract the dust.  
  
Essentially, our method to subtract the dust emission looks for a parameter,   
$\alpha(\nu)$, which depends on the frequency and is such that the difference  
\begin{equation}  
\xi(\nu, \vec{x}) = M(\nu, \vec{x}) - \alpha(\nu) M(857, \vec{x})  
\end{equation}  
has a minimum dust contribution in terms of the variance.   
$M(\nu, \vec{x})$ denotes the Planck map at frequency  
$\nu$ to which we want to   
subtract the dust, $M(857, \vec{x})$ is the map at 857 GHz (dust map)   
and $\xi(\nu, \vec{x})$ is the new map at frequency $\nu$  
after dust subtraction.  
Then, the $\alpha(\nu)$ parameter is determined by imposing that the variance   
of the residual map $\xi(\nu, \vec{x})$ must be a minimum.   
If we write down the expression for this variance:  
\begin{equation}  
{\sigma(\nu)^2}_R \propto \int \xi(\nu, \vec{x})^2 d\vec{x} =  
\int (M(\nu, \vec{x}) - \alpha(\nu) M(857, \vec{x}))^2 d\vec{x}   
\end{equation}  
and now we require that the derivative of ${\sigma(\nu)^2}_R$ with respect to   
$\alpha(\nu)$ must be $0$, we finally find:  
\begin{equation}  
\alpha(\nu) = \frac{\int M(\nu, \vec{x})M(857, \vec{x}) d\vec{x} }  
{\int M(857, \vec{x})^2 d\vec{x} }  
\end{equation}  
where $\int d\vec{x}$ refers to an integral in real space over all the pixels   
of the map. At the end, what we obtain is a value for $\alpha(\nu)$ which is   
different for each frequency. The values of $\alpha(\nu)$  
should approach the real frequency dependence of the dust at these frequencies.  
It is important to remark that by subtracting the dust with this method   
we are not perfectly subtracting this component. This method is assuming   
that the dust has the same frequency dependence in the analyzed sky patch   
which is a good assumption for small regions of the sky.   
However, as it will be noted in section \ref{section_application},    
this approach has been proved to be successful with our simulations   
where we have really considered a varying spectral index and grain  
temperature for the dust emission. We have computed   
the spatial distribution of the spectral index on our simulated dust maps.   
We found that is nearly constant with a dispersion of about $3\%$. \\  
We would like to note that our method to subtract the dust is   
nothing less but a Wiener Filter since we are looking for a linear operation on   
the maps such that the variance of the residual is minimum. This is just the   
general definition for Wiener filter (e.g Zaroubi et al. 1995). \\  
   
\noindent  
{\bf CMB.}  
  
\noindent  
CMB is the last component that can be subtracted easily.  
Since the $\Delta T/T$ distortion in the CMB map is independent   
of the frequency, a good approach for the contribution of the CMB   
in each of the channels can be obtained by filtering a given CMB-dominated   
channel to the resolution of the other channels.  
The selection of the optimal channel to be filtered   
is easy if we want to increase the S/N ratio of the SZ effect. The channel   
at 217 is expected to have a negligible contribution to the thermal SZ effect   
while the CMB component dominates over the other ones.   
Furthermore, this channel has a resolution (FWHM = 5.5 arcmin) very   
close to the best Planck resolution (FWHM = 5 arcmin).   
Moreover, this channel has a low noise level (although it is not   
the best channel in terms of the noise level).  
  
Therefore our first step in {\it cleaning} the maps of the CMB contribution   
is to filter the high frequency channels with FWHM = 5 arcmin to   
the resolution of the channel at 217 GHz   
(FWHM = 5.5 arcmin) (FWHM-Filter = $\sqrt{5.5^2 - 5^2}$) where we have assumed   
that the beam can be well described by a Gaussian.   
Then, the channel at 217 GHz is subtracted from the filtered channels.   
We repeat the process with the channels at frequencies below 217 GHz but in   
this case we have to filter the 217 GHz channel to the resolution of the   
other channels since in this case the channels below 217 GHz have   
a FWHM larger than 5.5 arcmin.\\

\noindent  
It is important to keep the previous order (PS, dust and CMB). Point sources   
should be extracted first (at least in the channel at 857 GHz) since the   
frequency dependence of the point sources will be in general different to   
the frequency dependence of the dust in our Galaxy. Therefore if we subtract   
the 857 GHz map including the point sources in that map we are assuming that   
these point sources have the frequency dependence of our Galaxy, which is wrong.  
Then we should subtract the dust prior to the subtraction of the   
CMB (217 GHZ map) since we need to reduce the dust contribution in the   
217 GHz map before subtracting this channel from the others.  
  
  
We have to point out that,  
in the process of subtracting the components,    
we have increased the S/N ratio for the SZE but we have also increased   
the noise level.  
First, when subtracting the dust, we are adding the noise level of the   
857 GHz map times the constant $\alpha(\nu)$ to the other maps.   
This process does not increase the noise level   
very much in the channels below 300 GHz. At these frequencies the CMB    
contribution starts to dominate over the thermal dust   
and the value of $\alpha(\nu)$ is, therefore, very small. Hence,    
the noise contribution is small as well. 
  
Since the S/N ratio of the dust in the 857 GHz map is high and the value  
$\alpha(\nu)$ is small at these low frequencies, then, the dust  
is subtracted without increasing the noise level too much in the low   
frequency channels. Furthermore, since the 857 GHz channel must be filtered,   
this process decreases the noise level even more.  
  
The increase in the noise level due to the CMB subtraction   
is more important in those channels where the resolution is closer to the   
one at 217 GHz (143, 353, and 545 GHz channels) and is less   
important in the other channels. This is because we have to filter the   
maps to the resolution of the lowest resolution channel. Hence, if the filter   
has a large FWHM (FWHM-Filter = $(FWHM_{\nu}^2 - FWHM_{217}^2)^{1/2}$),   
then the noise is smoothed over the large scale of the  
filter. As a result, some channels (around 217 GHz) have increased their    
noise level significantly while others did not change the noise level too much.   
  
\subsection{The Bayesian approach}\label{section_search_engine}  
\begin{figure}  
   \begin{center}  
   \epsfxsize=8.cm   
   \begin{minipage}{\epsfxsize}\epsffile{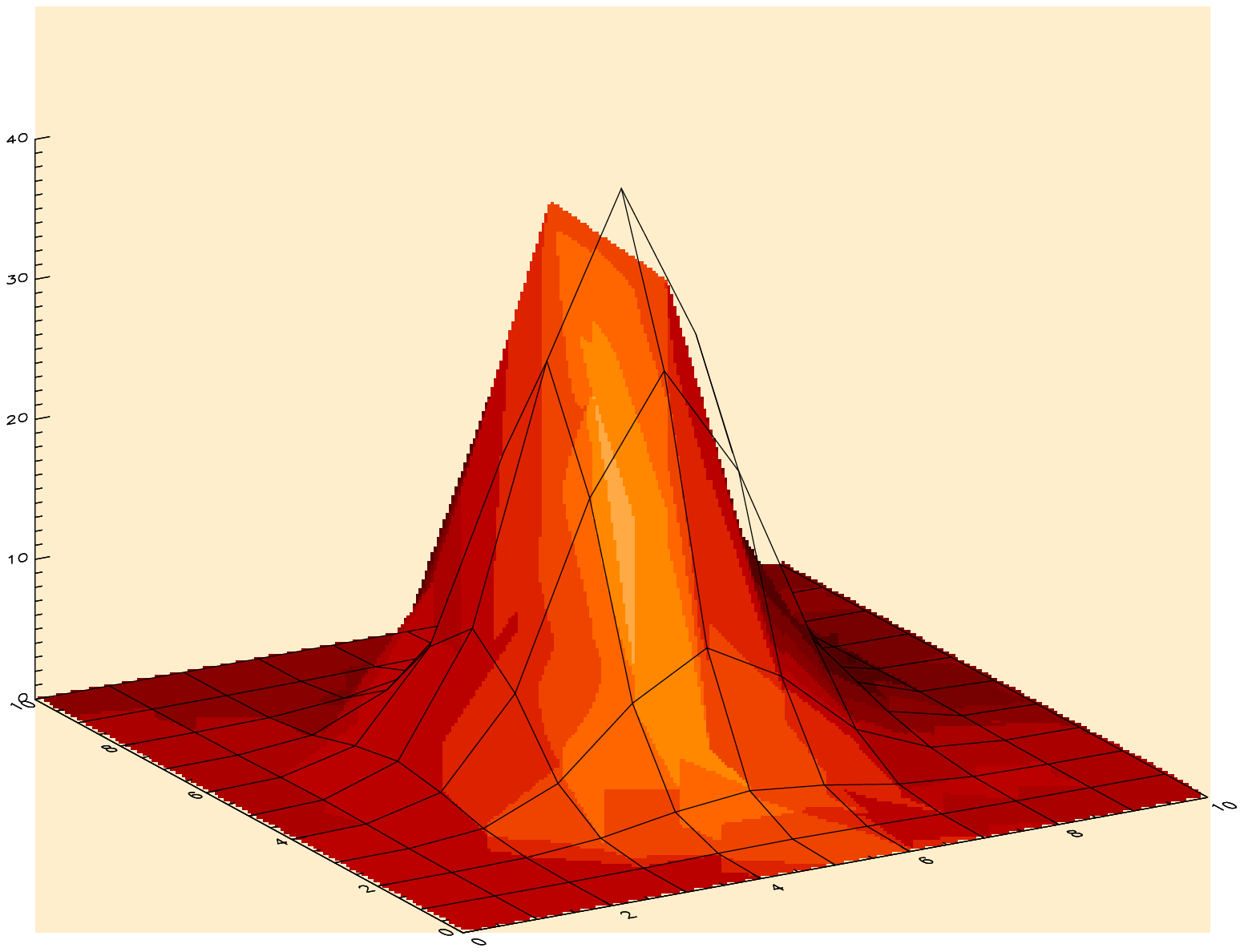}\end{minipage}  
   \caption{\label{fig_Prob_yc}  
            Probability distribution of the $y_c$ Fourier coefficients   
            (coefficients on the ring with $|\vec{k}| = 60$) for a   
            simulated SZE map. The histogram is made   
            in the Real-Imaginary plane ($X$ axis is the Real part of the Fourier   
            coefficient, $Y$ axis is the Imaginary part and $Z$ axis is the number of   
            Fourier coefficients with Real and Imaginary parts falling in   
            the bin $(X,Y)$). Those coefficients with Real   
            and Imaginary parts equal to 0 are in the center of the plane.   
            The grid shows the pdf given by expression (\ref{eq_Pyc})  
            with $P_{y_c}$ the power spectrum  of the simulated map at $k = 60$.  
            }  
   \end{center}  
\end{figure}  
  
After the previous steps have been applied, we end up with maps where   
some point sources have been (partially)   
removed, the dust and CMB contributions have been reduced and the noise   
level has been increased (in the high frequency channels more than in the   
low frequency ones).   
The resulting maps are dominated by the residuals of the point source   
subtraction (low frequencies) and noise (all frequencies)    
while the SZE contribution appears to be more important at  
intermediate frequencies.   
Since we are only interested on the SZ effect and not on the residuals nor   
the noise we can consider them as a {\it net} residual, $\chi(\nu, \vec{x})$.   
Then the data, $d(\nu, \vec{x})$ (Eq. \ref{eqn_DeltaT}, hereinafter  
we write $d(\nu, \vec{x})$, instead of $\Delta T/T(\nu, \vec{x})$ for  
simplicity), can be rewritten as:   
\begin{eqnarray}  
d(\nu, \vec{x}) &  = & SZ(\nu, \vec{x}) + \chi(\nu, \vec{x}) \\  
&  = & A(\nu)\otimes(f(\nu) y_c(\vec{x})) + \chi(\nu, \vec{x})  
\label{eq_hypo_RealSpace}  
\end{eqnarray}   
where $f(\nu)$ is the known frequency dependence of the thermal SZ effect   
(see Fig. \ref{fig_fx}), $y_c(\vec{x})$ is the Compton parameter we want to   
determine and $\chi(\nu, \vec{x})$ includes the residuals (point source,    
galactic components and CMB) and the instrumental noise.   
Due to the antenna convolution, $A(\nu)\otimes$, it is easier to work   
in Fourier space where the modes can be solved independently (provided   
the data is a homogeneous and isotropic field).   
Therefore, the previous equation should be rather expressed in Fourier space    
\begin{equation}   
d_{\nu}(\vec{k}) = A_{\nu}(\vec{k}) f_{\nu} y_c(\vec{k}) + \chi_{\nu}(\vec{k})  
\label{eq_hypo}  
\end{equation}   
where each mode can now be solved independently.  
We have adopted the {\it flat space} approach which is good enough    
when the scale of the map is $<< 1$ rad.   
Now we need some {\it engine} to search for the Fourier modes of the Compton parameter    
map which best resemble the real Fourier modes of the SZE in the    
data. We will use the Bayes theorem as such an engine:   
\begin{equation}  
P(y_c/d) \propto P(y_c)P(d/y_c)  
\end{equation}  
where the only thing we need to do is to define the prior $P(y_c)$   
and the likelihood of the data $P(d/y_c)$ and then look for the values of   
$y_c$ that maximize the posterior probability. \\

\noindent  
{\it The prior.}  
  
\noindent  
The prior should account for the probability of having a   
given Compton parameter (in Fourier space).   
Since there is not enough real data to know what a real SZ map looks,   
this probability should be obtained from simulations of the SZ.   
We found that, in Fourier space,   
the probability of having a given $y_c$ at each $k-$mode is close   
to a Gaussian of the form:  
\begin{equation}  
P(y_c) \propto exp(- {|y_c|}^2/P_{y_c}).  
\label{eq_Pyc}  
\end{equation}  
where  $P_{y_c}$ is the power spectrum of the SZ map, at each  
$k-$mode.  
  
We assume that the probability depends on $y_c$ through its module.   
This can be done due to the {\it almost} symmetric shape of the pdf   
of $y_c$ (see Fig.\ref{fig_Prob_yc}).   
The real pdf is not exactly symmetric as can be appreciated in the figure.   
This means that the real pdf (in Fourier space) is not exactly a Gaussian   
which is in accordance with the non-Gaussian nature of the SZE in real space.   
However, the pdf of the modes of the SZE map in Fourier space (although not exactly Gaussian)   
can be approached by a Gaussian much better than the pdf of the SZE map in real space. \\  
We found that the previous probability (equation \ref{eq_Pyc}) makes a reasonably   
good fit to the probability distribution of the $y_c$ at intermediate and   
large $\vec{k}$-modes (see Fig.\ref{fig_Prob_yc}).   
At small $\vec{k}$-modes, Eq.(\ref{eq_Pyc}) is still a reasonable    
approach although the small number of $y_c$ coefficients at these low $\vec{k}$-modes   
makes it difficult to have a (statistically) good enough probability   
distribution for $y_c$.  
Nevertheless, the interesting information containing the cluster   
contribution to the SZ effect is at intermediate $\vec{k}$-modes where Eq.(\ref{eq_Pyc})   
provides a reasonable approximation for the prior. The drawback of this approach is that we need to   
assume a given power spectrum for the SZ effect, $P_{y_c}$, in order to define the prior.\\  
  
\begin{figure}  
   \begin{center}  
   \epsfxsize=8.cm   
   \begin{minipage}{\epsfxsize}\epsffile{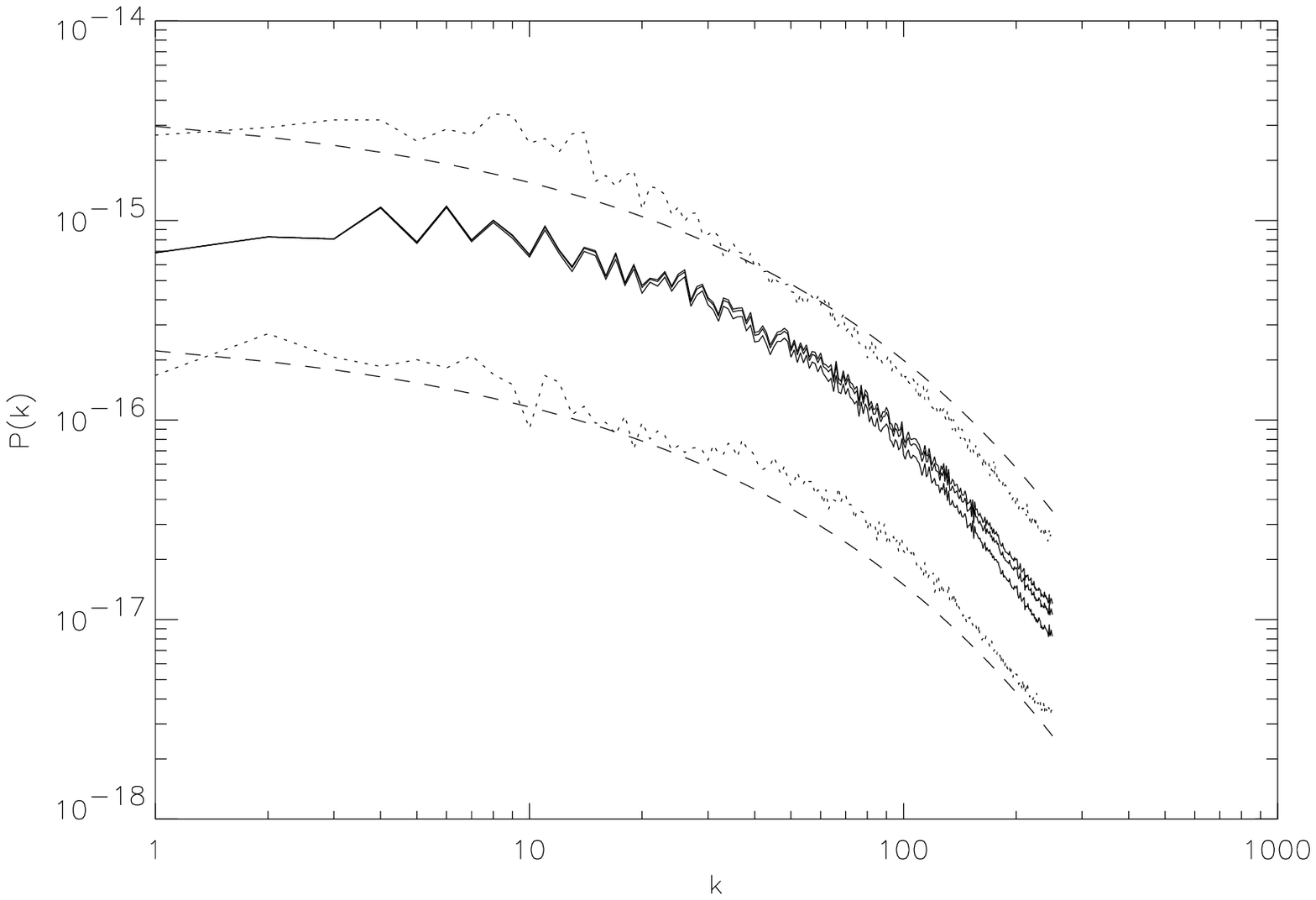}\end{minipage}  
   \caption{\label{fig_Pk_yc}  
            Power spectrum ($P_{yc}(k)$) for various simulations of the SZE. The three solid lines    
            are the power spectrums of three simulations with $r_v/r_c = 5, 10, 15$ and   
            $\sigma_8 = 0.8$. Dotted lines are two simulations with  $r_v/r_c = 10$   
            and $\sigma_8 = 0.7$ (down),  $\sigma_8 = 0.9$ (up).  
            The two dashed lines are the power spectra assumed in the prior in this work.   
            The upper dashed line corresponds to model A and the bottom dashed line corresponds   
            to model B.   
            }  
   \end{center}  
\end{figure}  
  
  
  
As shown in previous works (e.g Komatsu \& Seljak 2002), the power   
spectrum of the SZE is particularly sensitive to the value of $\sigma _8$.   
Our simulations also show this strong dependence. In Fig. \ref{fig_Pk_yc} we plot the   
power spectra for various simulations where we only change the value of $\sigma _8$   
and the internal structure of the clusters.   
The three solid lines are for $\sigma _8 = 0.8$   
and the dotted lines are for $\sigma _8 = 0.9$ (up) and   
$\sigma _8 = 0.7$ (down). The three solid lines show the change in the power spectra   
when we change the internal structure of the clusters. This is parametrized by a   
$\beta$-model ($\beta = 2/3$) with one free parameter, $r_v/r_c$ which is the ratio   
of the virial radius to the core radius.   
The different solid lines are for the cases  $r_v/r_c = 5, 10, 15$ and as it can be seen,   
the dependence on the internal structure is very weak. In the simulation of the SZE used to test   
the method we have assumed that $\sigma _8 = 0.8$ and $r_v/r_c = 10$ (central solid line).  
Hereafter, we will refer to this model as the 'true power spectrum'.   
  
Although the power spectrum of the SZE depends on the cosmological model (population)   
as well as on the internal structure of the individual clusters, a  
qualitative behaviour can be deduced straightforwardly: if we assume  
that the clusters are randomly distributed (following a Poissonian  
distribution) in the sky, the spectrum at low  $\vec{k}$-modes must be  
flat; on the other hand, at high $\vec{k}$-modes the power spectrum 
must go down, following the $\beta$-model profile.  
  
Hence, the shape of the power spectrum is rather stable but its amplitude   
strongly depends on the value of $\sigma_8$.   
In order to check the dependence of our method to the particular choice of the power spectrum,   
we will compare the results using different power spectra.  
The main results will be presented using the power spectrum   
shown as the upper dashed line in Fig. \ref{fig_Pk_yc} (model A). 
These results will be compared with the ones obtained using the true power spectrum   
(middle solid line) and the bottom dashed line (model B).   
  
The power spectrum shape for all the cases is an exponential of the form:  
\begin{equation}  
P_{y_c}(k) = C exp(-r \sqrt{k})  
\end{equation}  
This simple form of the power spectrum follows, more or less, the real   
power spectrum of our simulations (where the profile of the clusters  
is given by a $\beta$-model with $\beta = 2/3$) but is not exactly the same.  
We do not take the real power spectrum of our simulations because in a real situation   
this is not going to be known.   
The particular case shown in Fig. \ref{fig_Pk_yc} correspond to the parameters,   
$C=4.0 \times 10^{-15}, 3.0 \times 10^{-16}$ and $r = 0.3$ for models 
A and B respectively.  
The range in the normalization ($C \in [3.0 \times 10^{-16}, 4.0 \times 10^{-15}]$) corresponds   
more or less to a range in $\sigma_8 \in [0.7,0.9]$ which is consistent with the most recent   
constraints in this parameter (see e.g  Lahav et al. 2002 and references therein)  
  
The choice of an exponential form for models A and B is a compromise between taking the real power   
spectrum of the simulated SZE  and a different one which could have been obtained   
through a specific model.   
The specific shape of the power spectrum is not critical, but it is 
important that the chosen one does   not have much power at high $k$'s 
in order to suppress the deconvolution of the noise. This behaviour   
is easily achieved when the power spectrum is of exponential form.\\

\noindent  
{\it The likelihood.}  
  
\noindent  
If we assume as our hypothesis Eq.(\ref{eq_hypo}), then    
the residual $\chi_{\nu}(\vec{k})$ can be expressed as:  
  
\begin{equation}  
\chi_{\nu}(\vec{k}) = d_{\nu}(\vec{k}) - A_{\nu}(\vec{k}) f_{\nu} y_c(\vec{k})  
\label{eq_residual}   
\end{equation}  
or in vectorial form (each vectorial component being a different frequency):  
\begin{equation}  
\vec{\chi}(\vec{k}) = \vec{d}(\vec{k}) - \vec{R}(\vec{k}) y_c(\vec{k})  
\label{eq_residual}   
\end{equation}  
where the term $\vec{R}(\vec{k})$ is the response vector which has as many   
components as frequency maps considered. Each one of the components of the   
response vector is just:  
\begin{equation}  
R_{\nu}(\vec{k}) = A_{\nu}(\vec{k}) f_{\nu}.  
\end{equation}  
Again, each one of the components of the data vector $\vec{d}(\vec{k})$ is  
the Fourier component of the data (i.e. original data minus the estimates of point sources,   
dust and CMB) in the $\vec{k}$-mode and at frequency $\nu$. $\vec{\chi}(\vec{k})$   
contains the noise, the non-subtracted point sources, the free-free   
and synchrotron emissions, and the residual due to the subtraction of the thermal   
dust emission, the CMB and point sources.   
Although some of these   
components are certainly non-Gaussian, the fact is that this residual   
is clearly dominated by the noise (see Fig. \ref{fig_dust_217_substr} below)  
which can be very well approximated by a Gaussian. Therefore we will consider the   
hypothesis that the residual $\chi_{\nu}(\vec{k})$ can be well described by   
a Gaussian variable:   
\begin{equation}  
P(\vec{\chi_{\nu}}) \propto exp(-\vec{\chi_{\nu}} C^{-1}\vec{\chi_{\nu}}^{\dag})  
\label{eq_likelihood}  
\end{equation}  
where $C^{-1}$ is the inverse of the correlation matrix of the residual. This   
correlation matrix is not necessarily diagonal (and in fact it is not) since   
there are some correlations in the residuals between the different   
frequency channels.\\  
  
We are looking for the Fourier coefficients of the Compton parameter map, $y_c$,   
which minimizes the residuals in Eq. (\ref{eq_residual}). Since the   
terms in $C^{-1}$ are independent of $y_c$, the minimum of the   
residuals gives the maximum of the probability in Eq.(\ref{eq_likelihood}).   
Therefore Eq.(\ref{eq_likelihood}) can be considered as the likelihood of   
the data.   
  
Rewriting the posterior probability (for a given $\vec{k}$-mode)    
in terms of the data and the Compton parameter we finally get:  
  
\begin{eqnarray}  
P(y_c/d) &\propto& P(y_c)P(d/y_c) \\ \nonumber   
&\propto& exp(- \frac{{y_c}^2}{P_{y_c}})exp(-\vec{\chi_{\nu}} C^{-1}\vec{\chi_{\nu}}^{\dag})  \\ \nonumber  
&=& exp(- [{y_c}^2/P_{y_c} + (\vec{d} - \vec{R} {y_c})C^{-1}(\vec{d} - \vec{R} {y_c})^{\dag}]) \\ \nonumber  
&=& exp(- [ {y_c}^2/P_{y_c} + \vec{d}C^{-1}\vec{d}^{\dag} \\ \nonumber   
&-& 2\vec{d}C^{-1}\vec{R}^{\dag}{y_c}  \\ \nonumber   
&+& \vec{R}C^{-1}\vec{R}^{\dag}{y_c}^2])  
\end{eqnarray}  
where the latest equality follows from the symmetry of $C^{-1}$. \\   
Since we are interested in the value of $y_c$ which maximizes this   
probability we can simply look for the minimum of the exponential part by   
deriving it with respect to   
$y_c$ and making the result equal to $0$. The Compton parameter can then   
be found by simply solving for that equation:  
\begin{equation}  
y_c = \frac{\vec{d}C^{-1}\vec{R}^{\dag}}{\vec{R}C^{-1}\vec{R}^{\dag} + P_{y_c}^{-1}}.  
\label{eq_Search_engine}  
\end{equation}  
Notice that this result is the same than the one obtained with the  
multifrequency Wiener filter for the Compton $y_c$ parameter. 
 
Then, after returning to real space, we end up with a map   
of Compton parameters. The previous equation is our   
particular {\it search engine}. We want to remark that all the terms   
given in that equation can be computed directly from the data   
(see below).   
The only term for which some assumptions need to be made is for the   
power spectrum of the SZE, $P_{y_c}$, although the particular   
election of one or another shape for this power spectrum does not have   
any significant effect on the final result provided that it   
obeys some basic rules which will be discussed later.

\section{Applying the method to simulated data: Results}  \label{section_application}  
  
We have applied the method explained in the previous section  
to the simulated Planck data described in section \ref{data_set}.  
In Fig. \ref{fig_10chan} we show the 10 simulated data sets. These   
simulations include all the relevant components (galactic and   
extra-galactic) as well as the noise (we have considered that the noise   
is uncorrelated). The high frequency channels are clearly dominated by  
dust emission above $\nu \approx 300$ GHz.   
Some point sources can clearly be observed in these high frequency  
maps. Below $\nu \approx 300$ GHz the CMB starts to dominate over the other components.   
The synchrotron, free-free and spinning dust emission only  
contribute to the lowest channels although their contribution is expected to be small.  
First, we perform the map cleaning (partial subtraction of point sources,   
thermal dust and CMB). Then we apply our Bayesian approach in order to  
estimate the emission due to the SZE.  

\begin{figure}  
   \begin{center}  
   \epsfysize=18.cm   
\caption{Planck channels after point source and dust subtraction.   
         Note the effect in the 353 GHz channel. It was, basically,  
         a mixture of CMB and dust emission (see  
         fig. \ref{fig_10chan}) and now it is dominated by the   
         CMB contribution.}  
   \label{fig_dust_substr}  
   \end{center}  
\end{figure}  
\begin{figure}  
   \begin{center}  
   \epsfysize=18.cm   
\caption{Planck channels after point source, dust and CMB  
         subtraction. Low frequency channels are dominated by the  
         point source residuals. Some clusters can be seen by eye as black   
         spots in the 70, 100 and 143 GHz channels and as bright sources in   
         the 353 GHz channel.}  
\label{fig_dust_217_substr}  
\end{center}  
\end{figure}

\subsection{Map cleaning}  
  
Point sources have been extracted using the Mexican-Hat wavelet as    
was explained in section \ref{cleaning}.   
Applying the method described in Vielva et al. (2001a), we are able to detect  
(and subtract) 215 point sources in the 857 GHz channel, 25 at 545 GHz,  
27 at 353 GHz, 18 at 143 GHz,  
18 at HFI 100 GHz channel, 15 in the LFI 100 GHz channel, 12 at 70  
GHz, 9 at 44 GHz and 5 at 30 GHz.  
The number of spurious detections is lower than $5\%$ and the mean  
error in the amplitude estimation is lower than $18\%$ for all the  
channels.\\  
None of the clusters was identified as a point source in our simulation   
since the MHW only detect sources above a certain flux limit which is above   
the flux of the clusters in our simulation.    
In a much larger area of the sky (e.g. full sky) there could be some   
bright clusters with a large flux. However these high flux clusters   
are extended sources which can be easily distinguished from the   
point sources.\\  
  
The dust subtraction has been performed following the procedure  
explained in section \ref{cleaning}. The 857 GHz channel was filtered with an   
appropriate Gaussian beam in order to degrade its resolution to the other's     
channels resolution. We show the result in fig. \ref{fig_dust_substr}.  
As can be seen from the figure, the dust subtraction works very well at   
545 GHz where no appreciable galactic structure is seen.   
Also the quality of the dust subtraction   
can be observed in the 353 GHz channel. This channel was, before dust subtraction,   
basically a mixture of CMB and dust. After dust subtraction, the CMB   
component dominates clearly over the other components.  
  
The last component we subtract is the CMB itself.   
To do this, we just subtract the 217 GHz channel   
where the expected thermal SZ signal is negligible. Again, we have to   
filter this map to the resolution of the other maps and in the case   
in which the resolution of the other maps is smaller than the resolution   
of the 217 GHz channel, we filter the former maps to the    
resolution of the 217 GHz channel. The result can be seen in   
Fig. \ref{fig_dust_217_substr}

The main features that can be appreciated in these maps are the residuals   
of the point source subtraction. This fact suggests that an improvement   
in the method used to subtract the point sources could be to reduce   
the background  level by first subtracting the dust   
and then the CMB (after a prior point source subtraction in the 857 GHz   
and 217 GHz channels).  
Also some clusters can now be seen {\it by eye} in the channels between 100 and   
143 GHz as black spots and in the channel at 353 GHz as bright  
sources.  
  
It is important to note that {\it before} point source   
subtraction, the point source contribution was more important at the   
high frequency channels. After point source subtraction, however,   
the residual of the point sources is more important at low frequencies   
as it is shown in Figure \ref{fig_dust_217_substr}. \\  
The point source residuals at low frequencies are not a problem   
for our method since at those frequencies the method {\it searches} for   
negative wells. On the contrary, point sources are bright peaks which,   
therefore, can not be confused with galaxy clusters. The situation   
is different if a bright point source lies in the position of a   
cluster. In this particular case the point source is a serious contaminant   
for the detection of the SZE signature. \\  
The dust and CMB subtraction will also leave a residual in the maps although   
both residuals are small.   
However, the search engine will do account for these contributions to the residual since   
the correlation matrix, $C$, will be computed from the {\it clean} data maps shown   
in Figure \ref{fig_dust_217_substr}. \\

\subsection{Bayesian search engine}\label{subsect_BSE}  
  
The Fourier transforms of the maps shown in Fig. \ref{fig_dust_217_substr}   
is what we consider as our data, $\vec{d}(\vec{k})$, in the  
Bayesian approach of Eq. (\ref{eq_Search_engine}).  
In order to solve the previous equation for the   
parameters $y_c$, we need to compute the inverse of the correlation   
matrix: $C^{-1}$. This is the correlation matrix of the residuals, $\vec{\chi}(\vec{k})$,   
which is a $8 \times 8$ symmetric matrix since we have used 8 different channels   
(we did not include in the analysis the maps at 217 and 857 GHz since they were   
not {\it clean}, i.e. no dust and/or CMB subtraction has been performed in these channels).   
However we do not know the residuals until we know the   
Compton parameter (see Eq. \ref{eq_residual}).   
We solve this by running the code a first time   
taking the correlation matrix of the residuals as the correlation matrix   
of the data vectors, $\vec{d}(\vec{k})$,    
and obtaining a first guess of the Compton parameters, $y_c$,    
from Eq. (\ref{eq_Search_engine}).   
With this guess we can now compute the residuals (Eq. \ref{eq_residual}),   
their correlation matrix and finally its inverse, $C^{-1}$.  
  
Now we are ready to solve for Eq.(\ref{eq_Search_engine}) with a    
good estimation of the correlation matrix, $C^{-1}$.   
After solving Eq.(\ref{eq_Search_engine}) mode by mode in Fourier space we   
can go back to real space by computing the inverse Fourier transform. \\  

%
\begin{figure*}  
   \begin{center}  
   \epsfxsize=18.cm   
\caption{Recovered map of Compton parameters (right) versus input map (left).   
Both maps have been filtered with a Gaussian antenna of fwhm = 7 arcmin.   
This result corresponds to model A. The other two cases, model B and true power spectrum,   
look similar.   
}  
   \label{fig_SZ_recov}  
   \end{center}  
\end{figure*}  
  
The recovered $y_c$ map is shown in Fig. \ref{fig_SZ_recov} where   
we compare the recovered and the input maps.   
The method returns clusters at small and large scales simultaneously although   
the largest scales are not very well recovered due mainly to the   
low surface brightness of those clusters.   
The method also recovers shapes which are non-symmetric since we did not use   
any symmetrical filter to increase the S/N ratio (as it is done in the   
mexican hat wavelet analysis or the multi-filter method, Herranz et al. 2002,    
for instance). \\  
  
Since we can assume a nearly arbitrary shape for the power spectrum in the prior (models A and B),   
we can recover a biased estimate of the Compton parameter map. However, a  
percentage of this bias (the one related with the bad normalization), can partially   
be corrected if we re-scale the recovered map in a convenient way. \\  
To correct for the wrong normalization of the power spectrum, we can apply a technique similar to the   
one we used to remove the dust. A wrong normalization in the assumed power spectrum over-predicts   
(or under-predicts) the normalization of the recovered Compton parameter map. We can consider this normalization   
as a free parameter, $b$, and determine that parameter by requiring that the variance of the data minus $b$ times   
the recovered Compton parameter map is minimum.  
We normalize the Compton parameter map in Fourier space by requiring that the  
{\it global weighted variance}, $G_{\sigma}$, is minimum. 
\begin{equation} 
G_{\sigma} = \sum_{\nu} \sum_k \frac{f(\nu)^2(d_{\nu}(k) - b \times y_c(k))^2}{P_{\nu}(k)}  
\label{eq_G} 
\end{equation} 
where the sum over $\nu$ extends to the different channels used in the analysis and the  
sum over $k$ extends to the Fourier modes. Since the recovered Compton parameter only  
contains useful information up to a certain $k$ ($k_{max} \approx 30$), in the sum on  
$k$ we only include the Fourier modes up to $k_{max} = 30$.  
Each channel is weighted by the factor $f(\nu)^2/P_{\nu}(k)$ which accounts for the relative  
signal-to-noise of that channel.  
By deriving equation \ref{eq_G} with respect to the bias factor, $b$, we can find an expression  
for $b$ which can be used to renormalize the Compton parameter map by just multiplying it for $b$. \\ 
By correcting the bias in this way, the final Compton parameter map shows a very weak dependence on   
the assumed power spectrum in the prior as we will see below.   
\begin{figure}  
   \begin{center}  
   \epsfxsize=8.cm   
   \begin{minipage}{\epsfxsize}  
\end{minipage}  
\caption{ This figure shows the recovered map (power spectrum model A) versus the   
          true one (convolved with 7 arcmin FWHM and mean value subtracted) pixel by pixel.   
          The cross-correlation between the true and recovered   
          maps renders a coefficient, $r = 0.58$. We also show the best fitting linear   
          model, $y = sx$,  (dashed line) which has a slope $s = 0.58$. For comparison   
          we also show the expected case of a perfect recovery, $s = 1$  (solid line).}  
   \label{fig_yc_True_vs_yc_Recov}  
   \end{center}  
\end{figure}  
In Figure \ref{fig_yc_True_vs_yc_Recov} we show an $y_c^T - y_c^R$ plot where   
we represent pixel by pixel the recovered map versus the true map.  
For this plot, both maps have been filtered with a Gaussian beam (FWHM = 7 arcmin)   
since the recovered map does not contain information below certain scale while   
the true map does. The choice of the FWHM of the filter (7 arcmin) is based on the   
fact that the recovered map is basically build up from the contribution of the channels   
between 100 and 353 GHz which have resolutions between 5.5 and 10 arcmin (the initial   
resolution of 5 arcmin in the 353 GHz channel was degraded to 5.5 arcmin during   
dust subtraction). \\  
We have computed the cross-correlation coefficient;  
\begin{equation}  
r = \frac{\left < y_c^R(x,y) y_c^T(x,y) \right>}{\sigma _R \sigma _T}  
\end{equation}  
between the recovered ($y_c^R$) and true ($y_c^T$) filtered maps. The constants   
$\sigma _R$, $\sigma _T$ are the dispersions of the recovered and true maps   
respectively. We found a value of $r = 0.58$.   
We also have computed the slope of the best fitting model $y = sx$ and we found a   
slope of $s = 0.58$ ($s = 0.59$ when we use the true power spectrum and $s = 0.57$   
for the case of the model B).   
This value can be compared with the result obtained with Maximum Entropy   
(Hobson et al. 1998). They found a value, $s = 0.55$.\\

The recovered map contains true clusters as well as spurious detections.   
The spurious detections form a Gaussian distribution with mean value 0.   
This `{\it background} of spurious detections is due to our   
wrong (although approximated) assumption of Gaussianity in the pdf of the Fourier modes.    
Obviously, those pixels in the recovered map with Compton parameter smaller   
than approximately 0 (it is not exactly 0 because the analyzed map  
has zero mean, see Fig.~ \ref{fig_histo_yc})  
can be interpreted as spurious since this parameter is, by definition,   
positive. However, there will be still many pixels in the recovered map with positive   
values not corresponding to real clusters.   
The true clusters should be detected in this map which has an  
approximately Gaussian noisy background.    
We have used the image package \small{SEXTRACTOR}  
\normalsize (Bertin \& Arnouts, 1996) to discriminate between the true clusters and the spurious background.   
Basically, \small{SEXTRACTOR} \normalsize selects regions with connected pixels   
above a given threshold. This threshold is expressed in terms of the dispersion   
of the map ($\sigma$).     
After applying \small{SEXTRACTOR} \normalsize to the previous recovered image, it returns   
45 detections at the $3 \sigma$ level (regions with more than 15 pixels   
connected above the $3 \sigma$ threshold).   
>From those detections 44 were real clusters and 1 was a spurious detection.    
If the detection limit is decreased to $2 \sigma$,  
\small{SEXTRACTOR} \normalsize returns 169 detections, 90 of   
such detections being spurious and 79 real.   
In the other two models (model B and true power spectrum), we found the same number of detections,  
44 real and one spurious ($3 \sigma$). 
\begin{figure}  
   \begin{center}  
   \epsfxsize=8.cm   
   \begin{minipage}{\epsfxsize}\epsffile{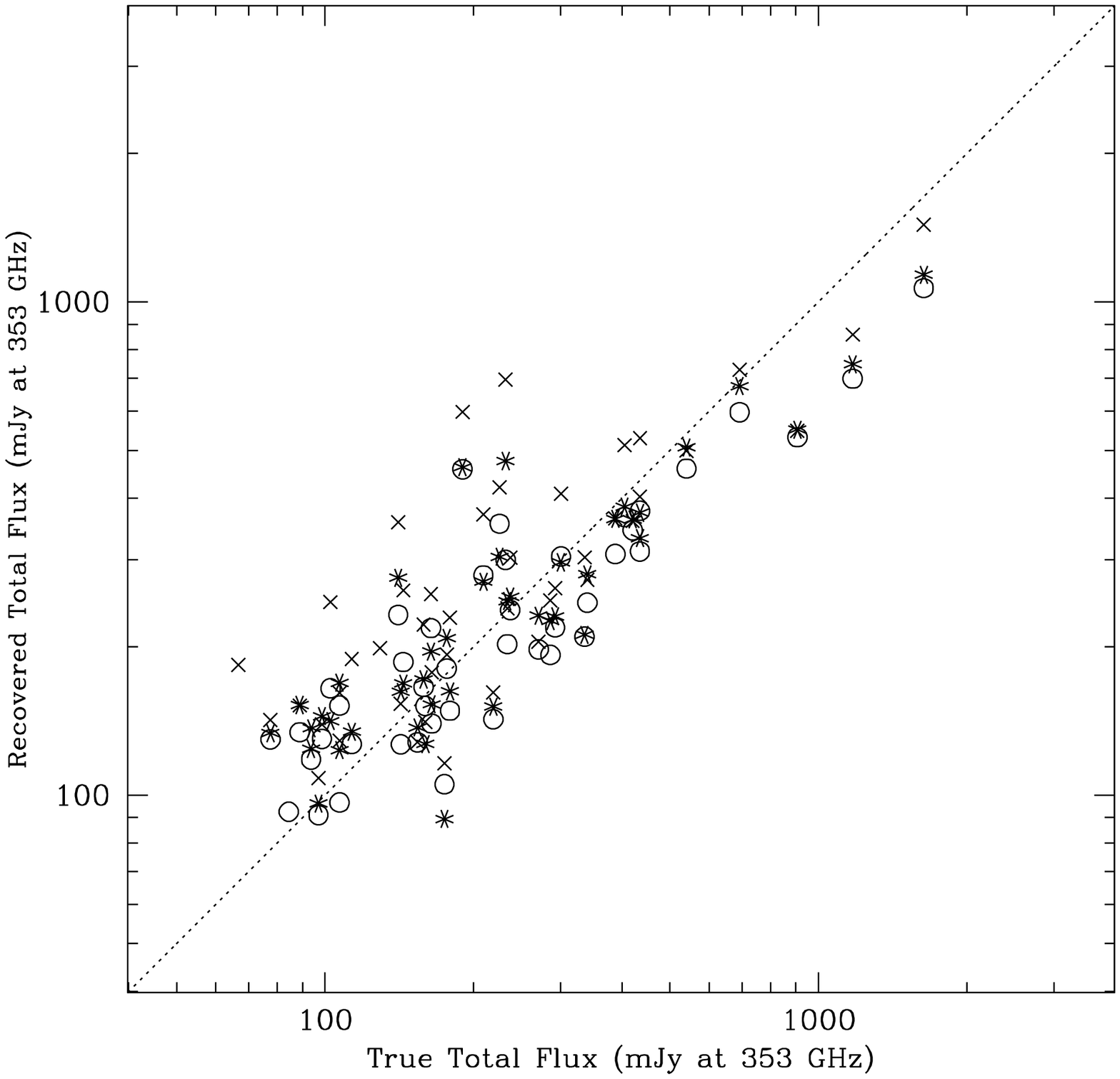}\end{minipage}  
   \caption{\label{fig_Flux_True_vs_Recov}  
             True total flux vs recovered total flux for the clusters returned   
             by \small{SEXTRACTOR} \small. The dotted line corresponds to the perfect   
             situation True Flux $=$ Recovered Flux. Circles are for model A, crosses for   
             model B and asterisks for the true power spectrum.   
             }  
   \end{center}  
\end{figure}  
\normalsize  
  
A large number of detections is important for cosmological studies. However, these   
detections should be useless if we do not have a good estimation of the flux of the clusters   
since this quantity is required to build the cluster number counts, $N(>S)$.  
In order to check for a possible bias of the recovered total flux in the SZE map we have   
compared the true and the recovered total fluxes as returned by  
\small{SEXTRACTOR} \normalsize.  
This result can be seen in Fig. \ref{fig_Flux_True_vs_Recov}.   
Our method recovers the real flux with no significant bias.    
The previous plots also show the good   
agreement among the results independently of the assumption made on the power spectrum of the SZE.   
Changing the power spectrum by one order of magnitude does not have a significant effect on the   
recovered map and the fluxes.\\   
  
\begin{figure}  
   \begin{center}  
   \epsfxsize=8.cm   
   \begin{minipage}{\epsfxsize}\epsffile{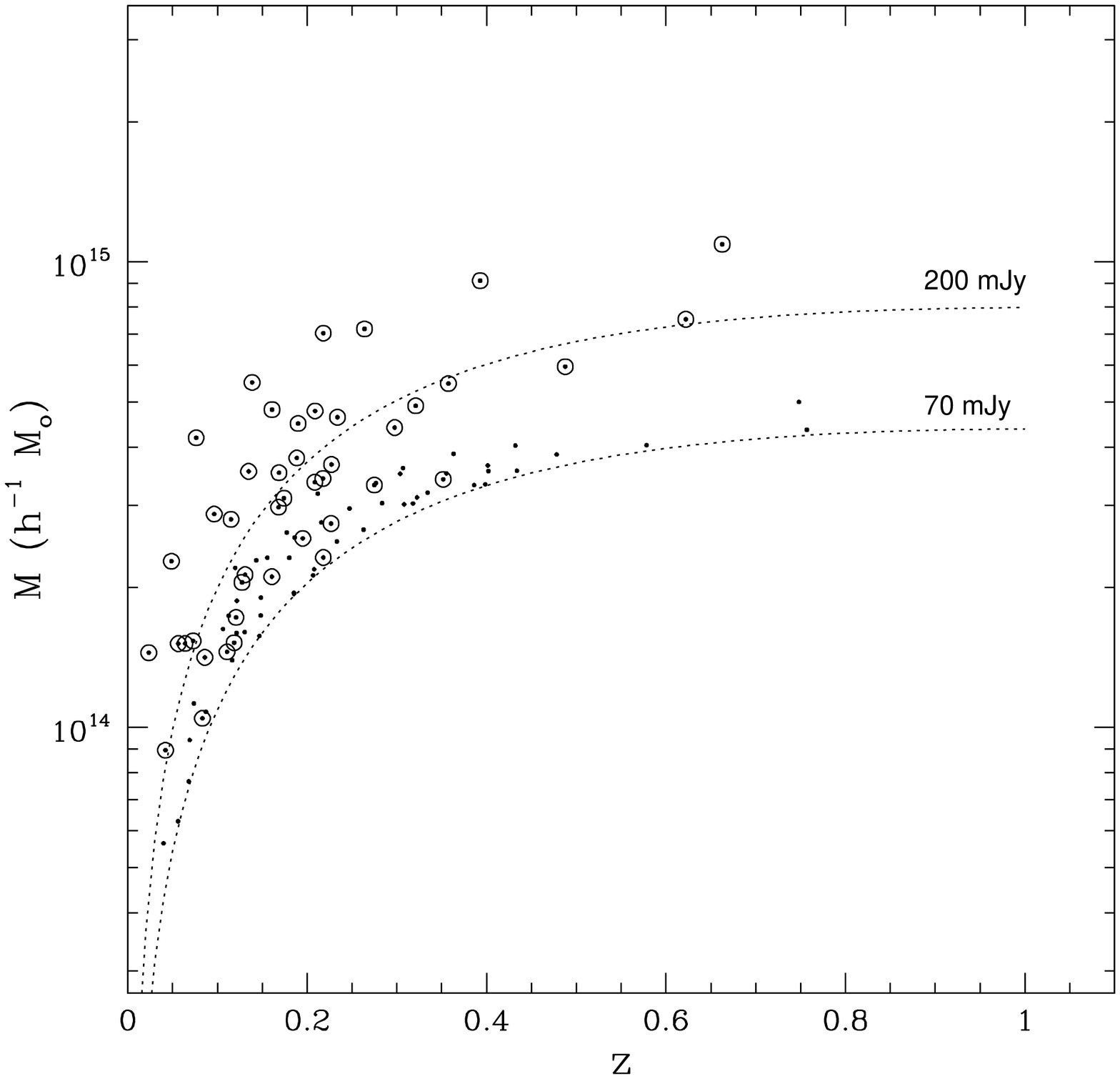}\end{minipage}  
   \caption{\label{fig_Mz}  
            Each point in this plot represents a cluster present in the simulation.   
            Clusters having a flux below 70 mJy (at 353 GHz) are not represented.   
            The clusters detected (15 pixels, 3$\sigma$) by  \small{SEXTRACTOR} \small   
            are marked with a big open circle around the small dots.   
            The two dotted lines are the selection functions   
            for a survey with limiting fluxes of 70 mJy (bottom) and 200 mJy (top) at 353 GHz.   
            Note that above 200 mJy, all the clusters in the simulation have been detected by   
            \small{SEXTRACTOR} \small. This implies that the Planck catalogue should be   
            complete above this flux.   
           }  
   \end{center}  
\end{figure}

Another parameter which is important for cosmological studies is the completeness of the   
recovered catalogue.   
The completeness level of the method is 100\% above fluxes $\approx 200$ mJy    
and drops quickly until fluxes $\approx$ 70 mJy (true flux) below which no cluster is detected   
(minimum flux detected was 77 mJy in model A, 67 mJy in the case of model B and 77 mJy in  
the case of the true power spectrum).    
This is illustrated in fig. \ref{fig_Mz} where we plot   
the detected clusters as a function of their mass and redshift. Also plotted are the clusters   
which have not been detected above the flux 70 mJy (all fluxes are given at 353 GHz).   
\normalsize  
  
  
In the other two cases for the power spectrum (model B and true power spectrum),  
the completeness level and selection functions  are very similar to the previous one.  
Therefore, the results are not very sensitive to the particular choice of the power spectrum in the   
prior, thus, allowing certain degree of freedom on its choice.   
However, in order to get satisfactory results, this power spectrum should satisfy a set of   
quite generic conditions which will be explained in the next section.\\  
  

\section{Conclusions and discussion}\label{section_discussion}  
  
We have presented a Bayesian and non-parametric method to detect the SZE signature in the Planck data.   
Our method only requires the knowledge of the frequency dependence of the SZE (which   
is well known) and the frequency dependence of the CMB (which is known to be constant).   
We also require that one of the channels must be completely dominated by dust emission.   
This is a well founded assumption for Planck where the 857 GHz channel is expected to   
be completely dominated by dust emission.   
We also need to assume a specific form for the power spectrum of the  
SZE component,  
but that form is clearly determined by the typical $\beta$-profile  
of the clusters.  
However, we have seen that the final result does not depend critically on the   
assumed SZE power spectrum in the prior. We have, therefore, certain degree of freedom on   
the choice of the power spectrum in the prior. However, there are several conditions, which   
this power spectrum should obey in order to get satisfactory results.  \\  
  
  
If the clusters are randomly distributed in the sky (following a  
Poissonian distribution) then the spectrum must be flat at low values  
of the $\vec{k}$-modes. Because clusters have a finite extension, the  
power spectrum must go down at large values of $\vec{k}$-modes. The specific  
fall-off depend on the cluster profile. For a $\beta$-model ($\beta =  
2/3$) the power spectrum can be described by the (Eq.~14).  
  
An exponential power spectrum, as the one assumed in this work, can obey both conditions.   
At high $k$-modes, it suppresses the noise deconvolution in a very effective way since it appears   
in the denominator of equation (\ref{eq_Search_engine}) as the inverse of an exponential.   
This kind of behaviour is also achieved by any power spectrum which is small at  
high $k$-modes.  
The normalization condition is satisfied when the inverse of the power spectrum in the    
low $k$-modes regime is not much larger than the term $\vec{R}C^{-1}\vec{R}^{\dag}$.   
This is a {\it normalization} constraint which can be satisfied by most of the cosmological   
models (with $\sigma_8$ within reasonable limits).\\  
The effect of the prior can be considered as a Fourier filter which suppresses   
the noise level while retaining the useful information at intermediate and low $\vec{k}-$modes. \\  
\begin{figure}  
   \begin{center}  
   \epsfxsize=8.cm   
   \begin{minipage}{\epsfxsize}\epsffile{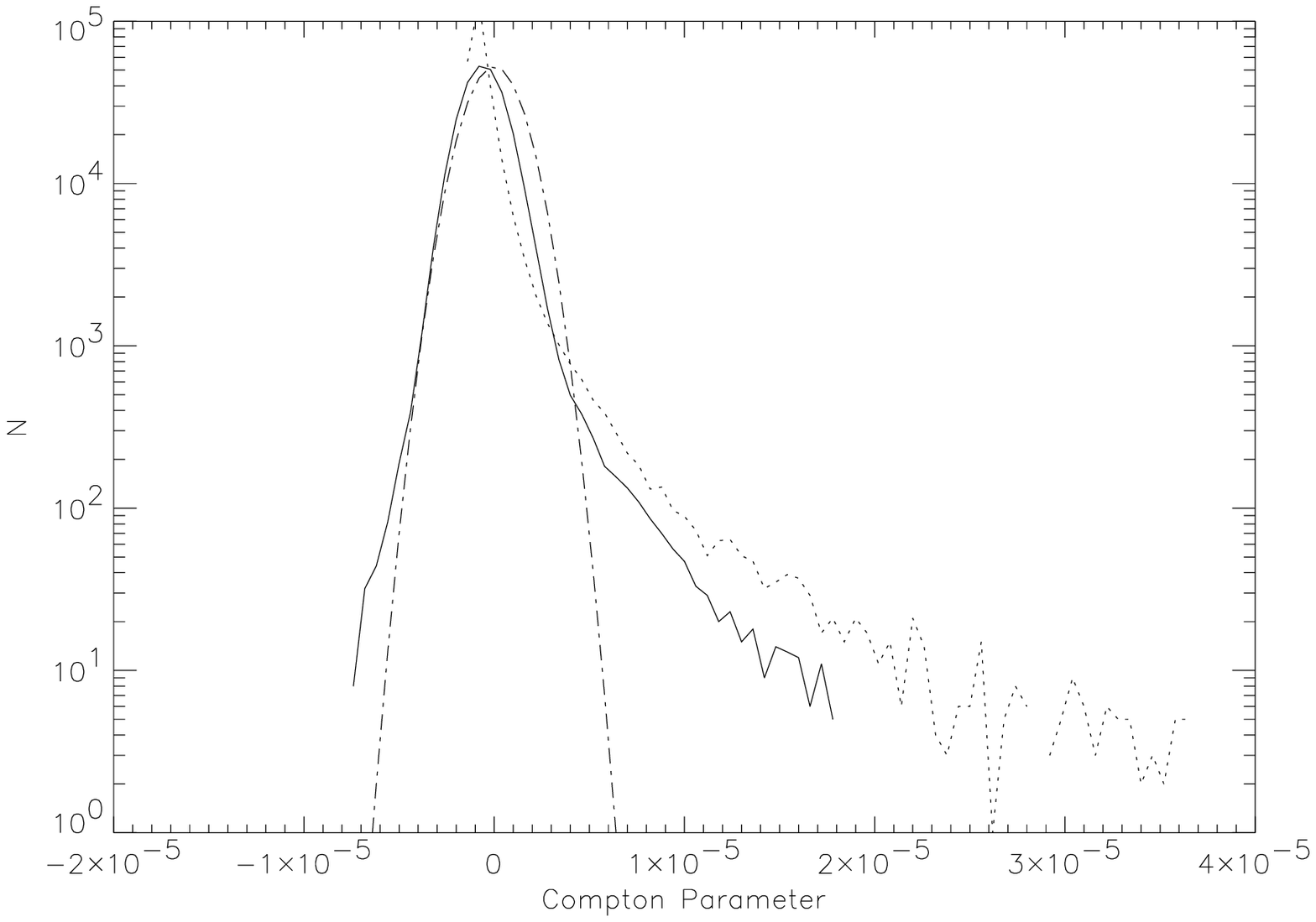}\end{minipage}  
   \caption{\label{fig_histo_yc}   
            The figure shows the pdf of the recovered map (solid line, model A). The   
            dot-dashed line is for a Gaussian with $\sigma =$ dispersion of   
            the recovered map. The dotted line is the pdf of the true map (mean value   
            subtracted). Both maps, true and recovered, have been convolved for this   
            plot with a Gaussian beam of FWHM $= 7$ arcmin.   
            The convolution is done because the recovered map   
            does not contain any information below 5 arcmin. We chose 7 arcmin since   
            the most relevant channels for the SZE have FWHM ranging from 5 arcmin   
            to 10 arcmin. The clusters returned after applying SEXTRACTOR are in   
            the positive tail of the pdf.  
           }  
   \end{center}  
\end{figure}  
%
  
The reader should note that equation \ref{eq_Pyc} is nothing but the expression   
for a Gaussian. That is, we are assuming that the Fourier coefficients of the SZE map   
are Gaussian. The inverse Fourier transform of a set of coefficients  
normally distributed is a map in the real space also normally distributed.  
However we know that the SZE is clearly non-Gaussian.  
In Fig. \ref{fig_histo_yc} we show the pdf of the recovered Compton parameter map (real space)   
and overlaid a Gaussian distribution (dot-dashed line).  
As can be seen, the recovered map is   
very close to a Gaussian distribution but it has a long positive non-Gaussian tail.   
When we apply \small{SEXTRACTOR}\normalsize, all the detections are in the non-Gaussian   
positive tail. Therefore we can consider our detections as the non-Gaussian part of   
an {\it almost} Gaussian pdf.  
The disagreement in the positive tail of the pdf when compared to the real pdf of  
the simulation can be explained by the bias in the recovery pixel-by-pixel  
(see figure \ref{fig_yc_True_vs_yc_Recov}). However, this is not in disagreement  
with the fact that the recovered flux is unbiased. Our method recovers the clusters  
with a larger apparent size but with a lower Compton parameter (see figure \ref{fig_SZ_recov}).  
Both effects compensate (after correcting for the bias, $b$)  
so the recovered flux shows no significant bias. 
Although in the method presented here we have assumed Gaussian  
distributions for the likelihood and the   
prior, the method can be applied in a more general (and maybe more realistic) way, considering    
that both, the likelihood and the prior, are non-Gaussian. \\  
  
We have seen that after applying \small{SEXTRACTOR}\normalsize  \  to our final recovered   
map we can detect $\approx 45$ clusters.   
A number of 45 detections in our small sky patch $(12.8^{\circ})^2$ means that we expect $\sim 11000$   
detections in all the sky ($\sim 9000$ in $4/5$ of the sky, i.e outside the Galactic plane).   
Such a large number of detections will allow detailed studies   
of the evolution of the cluster population which will have important cosmological consequences. \\  
We have seen that the method can reach limits   
of up to $\approx$ 70 mJy (at 353 GHz) although with a very low completeness level. In a   
previous paper (Diego et al. 2001b), we estimated the flux limit for the   
MEM method (Hobson et al. 1999). We found that    
limit to be $\approx 30$ mJy (353 GHz) although we do not know whether the   
MEM returns  an unbiased estimate of the total flux (at least   
the pixel-by-pixel recovery is biased as shown by the authors in   
Hobson et al. 1998) and we do not know the completeness level of MEM at this flux.  
An unbiased estimation of the flux is essential in order to build the $N(S)$   
curve (number of clusters with fluxes in  $[S, S+dS]$).  
Curves like this allow an independent determination of the   
cosmological parameters which should agree with the conclusions obtained   
from the CMB alone. Understanding of the selection function   
and completeness level of the survey is important   
since this information is needed to model the cluster number counts   
and its evolution. \\  
  
Our method provides an estimate of the total flux of the clusters   
which is consistent with no bias and almost independent of the assumed power  
spectrum.  
We also have seen that the catalogue of detected clusters is complete for those clusters   
with fluxes above 200 mJy. This flux defines the selection   
function of the catalogue which is needed in order to study the cluster   
number counts.  

\begin{figure}  
   \begin{center}  
   \epsfxsize=8.cm   
   \begin{minipage}{\epsfxsize}\epsffile{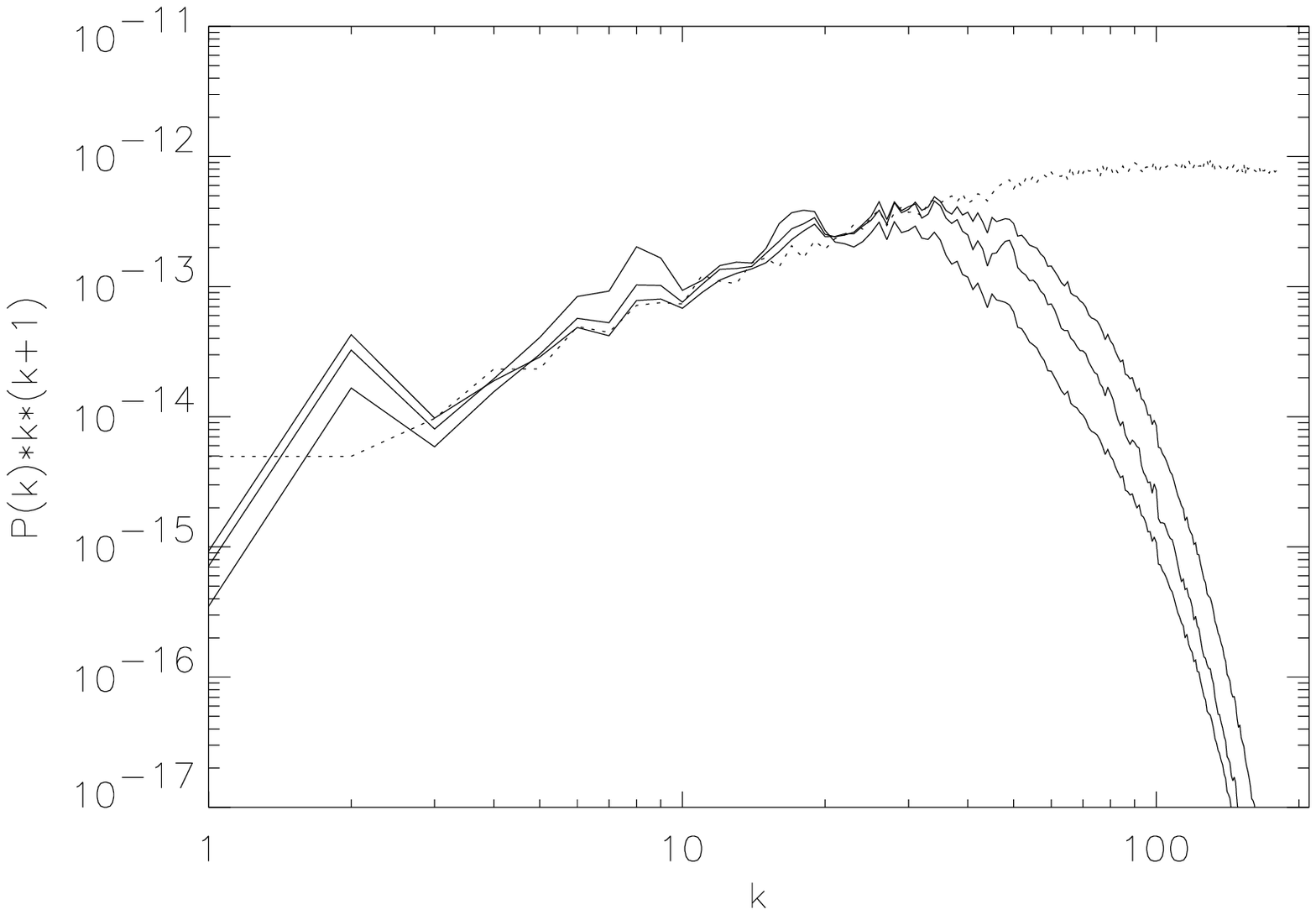}\end{minipage}  
   \caption{\label{fig_Pk_rec}  
             Power spectrum of the recovered Compton parameter map (solid lines) compared with the   
             real power spectrum (dotted line). The three solid lines correspond to the recovered   
             power spectrum for the three models (true power, model A and model B).   
             The estimate of the power spectrum given by our method could be used to define the   
             prior of the SZE in other methods.   
             }  
   \end{center}  
\end{figure}

As was shown in Diego et al. (2001b), the study of the number counts   
as a function of flux (and/or redshift) could produce strong constraints   
in the cosmological parameters ($\Omega$, $\sigma_8$, $\Gamma$).   
That work was based on a Planck cluster catalogue with a limiting   
flux of 30 mJy (353 GHz, MEM).   
For this flux limit we found that we should expect   
$\approx 30000$ clusters in 2/3 of the sky.   
If in fact, with Planck we are able to reach this limiting flux, the cosmological   
implications of such a large cluster catalogue would be very relevant.   
To reach this limit with a given separation method, one should be very   
cautious with the particular choice of the priors. In any case,  
our results are very robust and the information   
provided by this technique could be used to choose the best   
prior in other methods. For instance, the power spectrum (in  
particular its normalization) of the SZE is   
assumed to be known in most of the component separation algorithms. Our method could   
produce an estimate of this which could be used in these methods.   
As we show in Fig. \ref{fig_Pk_rec}, our method produces a rather accurate description of the   
real power spectrum up to $k \approx 30$ ($l \approx 900$). The three power spectra  
shown in  Fig. \ref{fig_Pk_rec}   
correspond to the three recovered power spectra for models A, B and true power spectrum.  
Our estimate of the power spectrum is very robust in the sense that it does not   
show any strong dependence on the assumed power spectrum in the prior (at least up to $k \approx 30$).   
Our estimation of the power spectrum could be used, for instance, to define the normalization of   
the power spectrum which is needed in other methods.    
As we suggested in the introduction, a successful   
method to perform the component separation should combine several   
methods. The technique presented in this paper could be just a part of   
the final method. \\

\section{Acknowledgments}  
This research has been supported by a Marie Curie Fellowship   
of the European Community programme {\it Improving the Human Research   
Potential and Socio-Economic knowledge} under   
contract number HPMF-CT-2000-00967.  
This work has been supported by the Spanish DGESIC Project    
PB98-0531-C02-01, FEDER Project 1FD97-1769-C04-01, the   
EU project INTAS-OPEN-97-1192, and the RTN of the EU project     
HPRN-CT-2000-00124. \\  
PV acknowledges support from a fellowship of Universidad de Cantabria.

  
  

  
\bsp  
\label{lastpage}  
\end{document}